\documentclass[pre,aps,twocolumn, floatfix,
superscriptaddress,showpacs]{revtex4-1}
\usepackage{amsmath,bm,graphicx}
\usepackage{amssymb}
\usepackage{amsfonts}
\newcommand{\pd}[2]{\frac{\partial #1}{\partial #2}}
\newcommand{\dd}[2]{\frac{d #1}{d #2}}

\begin{document}

\title{Collective oscillations in driven coagulation}

\author{Robin C. Ball}
\email{R.C.Ball@warwick.ac.uk}
\affiliation{Centre for Complexity Science, University of Warwick, Gibbet Hill
Road, Coventry CV4 7AL, UK}
\affiliation{Department of Physics, University of Warwick, Gibbet Hill Road,
Coventry CV4 7AL, UK}
\author{Colm Connaughton}
\email{connaughtonc@gmail.com}
\affiliation{Centre for Complexity Science, University of Warwick, Gibbet Hill
Road, Coventry CV4 7AL, UK}
\affiliation{Mathematics Institute, University of Warwick, Gibbet Hill Road,
Coventry CV4 7AL, UK}
\author{Peter P. Jones}
\email{P.P.Jones@warwick.ac.uk}
\affiliation{Centre for Complexity Science, University of Warwick, Gibbet Hill
Road, Coventry CV4 7AL, UK}
\author{R. Rajesh}
\email{rrajesh@imsc.res.in}
\affiliation{
Institute of Mathematical Sciences, CIT Campus, Taramani, Chennai-600113,
India}
\author{Oleg Zaboronski}
\email{O.V.Zaboronski@warwick.ac.uk}
\affiliation{Mathematics Institute, University of Warwick, Gibbet Hill Road,
Coventry CV4 7AL, UK}

\date{\today}

\begin{abstract}
We present a novel form of collective oscillatory behavior in the kinetics of 
irreversible coagulation with a constant input of
monomers and removal of large clusters. For a broad class of collision
rates, this system reaches a non-equilibrium stationary state at large times 
and the cluster size distribution tends to a 
universal form characterised by a constant flux of mass through the space of 
cluster sizes. Universality, in this context, means that the stationary
state becomes independent of the cut-off as the cut-off grows. 
This universality is lost, however, if the aggregation rate between large and
small 
clusters increases sufficiently steeply as a function of cluster sizes. We 
identify a transition to a regime in which the stationary 
state {\em vanishes} as the cut-off grows.  This non-universal
stationary state becomes unstable, however,  as the cut-off is increased
and undergoes a Hopf bifurcation. After this
bifurcation, the stationary kinetics are replaced by persistent and periodic 
collective 
oscillations. These oscillations carry pulses of mass through the space 
of cluster sizes. As a result, the average mass flux remains constant.
Furthermore, 
universality is partially restored in the sense that the scaling of the period 
and amplitude of oscillation is inherited from the 
dynamical scaling exponents of the universal regime. The implications of
this new type of long-time asymptotic behaviour for other driven 
non-equilibrium systems are discussed.
\end{abstract}

\pacs{82.40.Bj,82.40.Ck,83.80.Jx}
%\keywords{}
\maketitle

The statistical dynamics of irreversible coagulation have been studied for 
almost a century since the pioneering work of Smoluchowski on Brownian 
coagulation of spherical droplets. See \cite{LEY2003} for a modern review. 
It nevertheless remains an important branch of statistical physics.
This is in part due to its status as a paradigm of
non-equilibrium kinetics, but primarily due to its connections to variety of
important modern 
problems. We particularly highlight applications in cloud
physics \cite{FFS2002}, surface growth \cite{KA2011} and planetary
physics \cite{BBK2009}. In these examples, coagulation of clusters is
supplemented with a source (or effective source in the case of \cite{BBK2009}) 
of small clusters or ``monomers''. Such driven coagulation, in which monomers
are
supplied to the system at a constant rate, is the main
focus of this article. One may expect the kinetics of such
a system to become stationary for large times \cite{WHI1982} with the loss of
clusters
due to coagulation compensated by the supply of new clusters provided by the
input of monomers. We show below that this intuitive picture is not
always correct and demonstrate the possibility of a new and strikingly different
long
time behavior characterised by time-periodic oscillatory kinetics.

Before we begin, let us introduce a large mass cut-off, $M$. Above this size
clusters are removed from the system. Physically this could be literal removal
as in the case of large droplets preferentially precipitating out of a cloud, or
quenching of reactivity due, for example, to charge accumulation. Our primary
motivation for introducing it, however, is theoretical
and we shall focus on what happens as $M \to \infty$. The basic quantity of interest
is the
cluster size distribution denoted by $N_m(t)$. It is the average density
of clusters of mass $m$ at time $t$. Assuming that the system is statistically
homogeneous, $N_m(t)$
has no spatial dependence. We denote the coagulation rate between clusters (or
coagulation ``kernel'') by $K(m_1, m_2)$. Suppressing the $t$-dependence of
$N_m(t)$ for brevity, the mean-field kinetics satisfy Smoluchowski's equation:

\begin{eqnarray}
\label{eq-Smol}
\partial_t N_m &=& \frac{1}{2}\,\int_1^{m}dm_1 K(m_1,m-m_1) N_{m_1} N_{m-m_1}\\
 \nonumber &-& N_m  \int_1^{M-m}dm_1 K(m,m_1) N_{m_1} + J\ \delta(m-1)\\
\nonumber &-& D_M\left[ N_m\right]
\end{eqnarray}
where
\begin{equation}
\label{eq-dissipation}
D_M\left[ N_m\right] = N_m \int_{M-m}^{M} dm_1\, K(m,m_1)\, N_{m_1}
\end{equation}
removes clusters larger than $M$ and $J$ 
is the monomer injection rate. We study the family of kernels
\begin{equation}
\label{eq-modelKernel}
K(m_1,m_2) = {\small \frac{1}{2}}\left(m_1^\nu m_2^\mu + m_1^\mu m_2^\nu
\right),
\end{equation}
which includes many of the commonly studied models \cite{LEY2003}. 
Eq.~\eqref{eq-modelKernel} can also capture the asymptotics of
most physically relevant kernels. We mostly consider cases for which
$\mu+\nu<1$.
This avoids complications due to gelation \cite{LEY2003}. The stationary
solution of Eq.~\eqref{eq-Smol} without cut-off was
found in \cite{HAK1987}. It is a power law for large $m$: 
\begin{equation}
\label{eq-Cab}
N_m = \sqrt{\frac{J\,[1-(\nu-\mu)^2]\cos[\pi(\nu-\mu)/2]}{4\pi}} \,
m^{-\frac{\nu+\mu+3}{2}}.
\end{equation} 
The exponent $\frac{\mu+\nu+3}{2}$ implies a constant flux of mass through the
space of sizes, $m$. It is a standard example of a non-equilibrium stationary
state with a conserved current. From Eq.~\eqref{eq-Cab}, this stationary state
exists only if $\left|\nu-\mu\right|<1$, a fact which is true for any scale
invariant kernel \cite{CRZ2004}. One might ask what happens if
$\left|\nu-\mu\right|>1$? This can occur in practice. Examples include coagulation
of ice clusters in planetary rings \cite{BBK2009}, gravitational clustering
\cite{KON01} and droplet sedimentation in static
fluids \cite{PK1997}.

The fact that the constant flux stationary state only exists for a certain class
of kernels has long been appreciated in the theory of wave kinetics
\cite{ZLF92}. There, the constraint $\left|\nu-\mu\right|<1$ would be
interpreted in terms of universality. If one solves the stationary version of
Eq.~\eqref{eq-Smol} with finite cut-off, $M$, and studies the behavior as
$M\to\infty$ one finds that when $\left|\nu-\mu\right|<1$, the leading order
terms becomes independent of $M$ as $M\to\infty$. The stationary
state thus tends to the above universal form found in \cite{HAK1987}. If, on the
other hand, $\left|\nu-\mu\right|>1$, the stationary state is non-universal and
retains a 
dependence on $M$ as $M\to\infty$. This phenomenon is referred to as {\em
nonlocality} of interaction (in the mass space) in the sense that all masses
remain strongly coupled
to the largest and smallest masses in the system. By extension, the interactions
in the regime $\left|\nu-\mu\right|<1$ are termed {\em local} although this is a
rather
weak form of locality. The presence of a finite cut-off is essential to obtain a
stationary state in the nonlocal regime as discussed in \cite{KC2012}.

\begin{figure}
\includegraphics[width=\columnwidth]{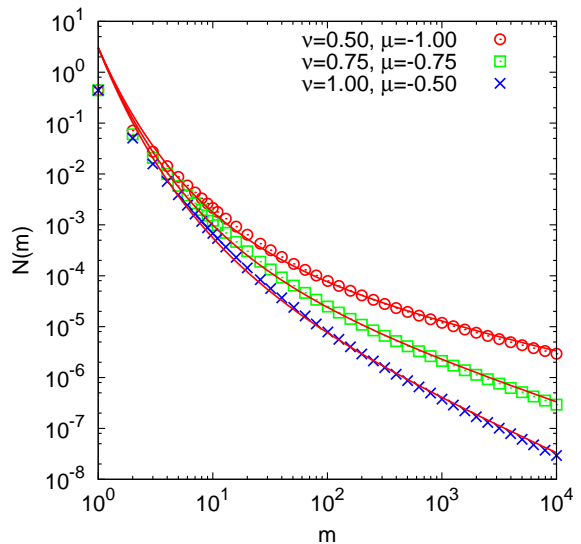}
\caption{\label{fig-nonlocalSS}Comparison of asymptotic approximation,
Eq.~(\ref{eq-nonlocalSS}) (solid lines) to
the true stationary state of Eq.~(\ref{eq-Smol})
(symbols) with the kernel given by Eq.~(\ref{eq-modelKernel}) for several values
of
$\nu$ and $\mu$ chosen in the nonlocal regime. The cut-off is $M=10^4$.}
\end{figure}

Almost nothing is presently known about the shape of $N_m$ in the nonlocal
regime. We developed an algorithm to compute the exact stationary solution of
the discrete version of Eq.~\eqref{eq-Smol} with cut-off by converting it into a
two-dimensional minimisation problem which can be easily solved numerically for
modest values of $M$. For details see Appendix \ref{app-generateStatDist}. Some typical
results are shown by the symbols in Fig.~\ref{fig-nonlocalSS}. It is clear that
the nonlocal stationary state is not a simple power law. To obtain some analytic
understanding, one possible way forward was outlined in \cite{HNS2008}. If
clusters of size $m$ grow primarily by interaction with clusters of mass $m_1
\ll m$, which is the essential feature of nonlocal interactions, one can Taylor
expand the righthand side of Eq.~\eqref{eq-Smol} and obtain an almost linear
equation for $N_m(t)$ \cite{HNS2008}. The
dominant terms in this equation are 
\begin{equation}
\label{eq-nonlocalSmol}
\pd{N_m}{t} = -D_{\mu+1}  \pd{}{m} \left[m^\nu N_m \right] - D_\nu\, N_m,
\end{equation}
where the $t$ dependence of $N_m$ has been suppressed and
\begin{eqnarray*}
&D_{\mu+1} = \int_1^\frac{m}{2} m_1^{\mu+1} N_{m_1} d m_1 & \to \int_1^M
m_1^{\mu+1} N_{m_1} d m_1,\\
&D_\nu = \int_m^M m_1^\nu N_{m_1} d m_1 & \to \int_1^M m_1^\nu N_{m_1} d m_1.
\end{eqnarray*}
Extension of the limits of integration of these latter integrals to $M$ and $1$
respectively is a further assumption which needs to be justified a-posteriori.
The self-consistent calculation detailed in \cite{BCSZ2011} for the case $\mu=0$
is easily extended to obtain the following stationary asymptotic solution of
Eq.~\eqref{eq-nonlocalSmol}
in the limit of large $M$:
\begin{equation}
\label{eq-nonlocalSS}
N_m^* \sim \sqrt{2\,\gamma\,J\,\log\left(M\right)}\, M^{-1}\,M^{m^{-\gamma}}
m^{-\nu} 
\end{equation}
where $\gamma=\nu-\mu-1$, adopting the convention that $\nu>\mu$ in
Eq.~\eqref{eq-modelKernel}. Detailed derivations of Eqs. \eqref{eq-nonlocalSmol}
and \eqref{eq-nonlocalSS} are provided in Appendices \ref{app-differentialApprox} and \ref{app-nonlocalSolution}.
Equation~\eqref{eq-nonlocalSS} approximates well the true stationary state as
indicated by the solid lines in Fig.~\ref{fig-nonlocalSS}. Note that there are
no adjustable parameters. A striking feature of Eq.~\eqref{eq-nonlocalSS} is
that the 
prefactor of the stationary state {\em vanishes} as $M\to\infty$ reflecting the
non-universality of the nonlocal regime. Similar behaviour was observed in the
instantaneous gelation regime in \cite{BCSZ2011} although there is no gelation
here.

\begin{figure}
\includegraphics[width=\columnwidth]{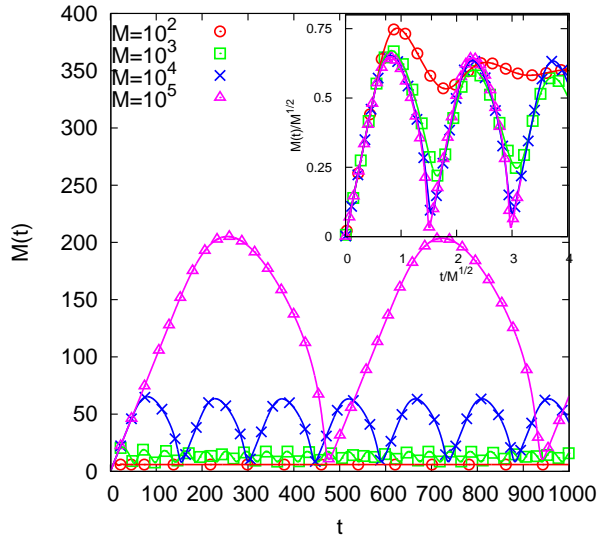}
\caption{\label{fig-period}
Main panel: Total mass vs time for different values
of $M$ with $\nu=-\mu=\frac{3}{2}$. Inset: Collapse obtained by rescaling the
data according to Eqs.~\eqref{periodScaling}.
}
\end{figure}

The vanishing of the stationary state in the limit $M\to\infty$ poses a
conceptual problem since it suggests the removal of the conduit linking the
source to the sink. In order to investigate how the conserved mass current is
carried in the nonlocal regime, we computed dynamical solutions of
Eq.~\eqref{eq-Smol} in the nonlocal regime using the numerical algorithm
developed in \cite{LEE2000,*LEE2001}. The results were surprising. For small
values of $M$, the numerical solution converged to the exact stationary
state as expected. Once $M$ exceeded a certain value, however, the numerical
solution never reached the stationary state. The typical behaviour of the total
mass as a function of time for different values of $M$ is shown in the main
panel of Fig.~\ref{fig-period} for the case $\nu=-\mu=\frac{3}{2}$. Stationarity
is reached only for smaller values of $M$. For larger $M$ we observe collective
oscillations which seem to persist indefinitely (we stopped the computation
after several hundred periods). The period and amplitude grow with $M$.

\begin{figure}
\includegraphics[width=\columnwidth]{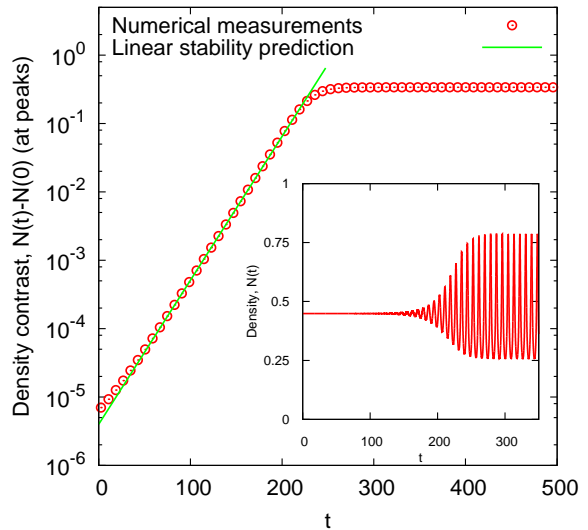}
\caption{\label{fig-instability} Numerical evolution of a perturbation of the
stationary state for $\nu=-\mu=\frac{3}{2}$ and $M=100$. Main panel: amplitude
of successive maxima of the perturbation (circles). The solid line is the
prediction of linear stability analysis. Inset: oscillations of the total
density. 
}
\end{figure}

\begin{figure}
\includegraphics[width=\columnwidth]{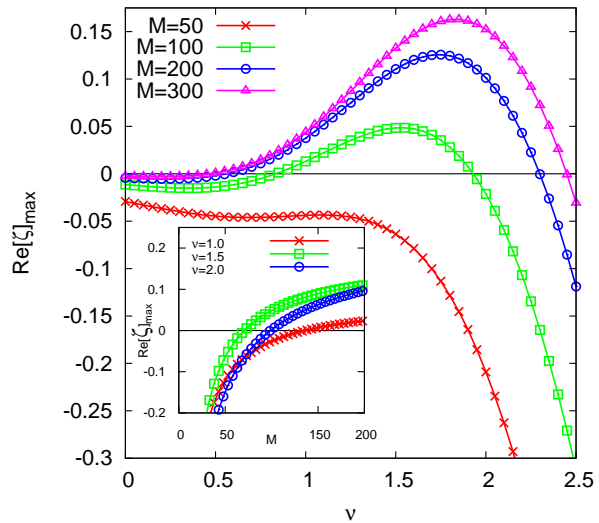}
\caption{\label{fig-bifurcation}
Main panel: $\mathrm{Re}[\zeta]_\mathrm{max}$ for kernels $\mu=-\nu$ plotted as
a function of $\nu$ for
different values of the cut-off, $M$. Inset: $\mathrm{Re}[\zeta]_\mathrm{max}$
as a function of $M$ for different values of $\nu$.
}
\end{figure}

\begin{figure}
\includegraphics[width=\columnwidth]{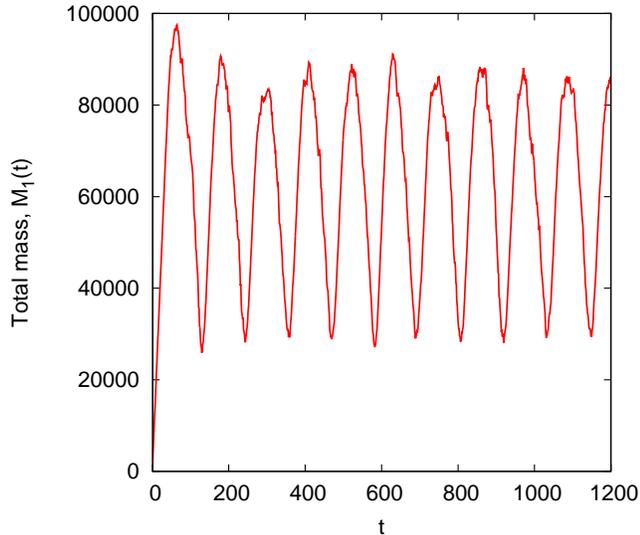}
\caption{\label{fig-MonteCarlo}
Total mass vs time in a Monte-Carlo simulation of the
Markus-Lushnikov model with a source and sink and kernel given by
Eq.~(\ref{eq-modelKernel}) with $\mu=-\nu=-0.95$ and $M=300$.  }
\end{figure}

The intriguing possibility thus arises that the stationary state becomes
unstable as $M$ increases. Our algorithm for computing the stationary state is
not dynamical and makes no distinction between stable and unstable fixed points.
We therefore input the exact stationary state as an initial condition for the
dynamical code and added a small perturbation. The results for the density are
shown in the inset of Fig.~\ref{fig-instability}. The perturbation grows to a
finite amplitude in a clear indication of instability. A lin-log plot of the
amplitude of the successive maxima of the perturbation, as shown in the main
panel of Fig.~\ref{fig-instability}, indicates exponential growth, a clear sign
of linear instability. We used {\em Mathematica} to compute the eigenvalue,
$\zeta_\mathrm{max}$, of the linearization of the discrete version of
Eq.~\eqref{eq-Smol} about the stationary state having maximum real part. This
analysis confirmed the instability. The growth rate agrees well with numerics
(see main panel of 
Fig.~\ref{fig-instability}). For fixed $\nu$ and $\mu$, the stationary state
undergoes a Hopf
bifurcation as $M$ is increased. The eigenvalue $\zeta_\mathrm{max}$ crosses the
imaginary axis at a critical value of $M$ (see inset of
Fig.~\ref{fig-bifurcation}) giving birth to a limit cycle and oscillatory
behavior.   The structure of the instability as a function of $\nu$ and $\mu$
for fixed $M$ is non-trivial as shown in the main panel of
Fig.~\ref{fig-bifurcation}. For fixed $M$, the stationary state becomes stable
again for sufficiently large values of $\nu$, a fact for which we have no
intuitive explanation at present. Such limit cycles appearing in mean-field
equations can be destroyed by noise\cite{MOB2010}. To check the robustness of
this phenenomenon, we performed Monte-Carlo simulations of the Markus-Lushnikov
model (see \cite{ALD1999}) with a source and sink of particles. Typical results
are shown in Fig.~\ref{fig-MonteCarlo}. Oscillations are clearly visible which
remain coherent in the presence of noise.

To understand nonlinear aspects of the instability such as the period and
amplitude of nonlinear oscillation we return to Eq.~\eqref{eq-Smol}. Each period
corresponds to a pulse of mass through the space of sizes. A movie provided
with the arxiv version of this paper illustrates these pulses. For details of the parameters 
see Appendix \ref{app-movie}. Each pulse almost resets the mass of the system
to zero as evident from the main panel of Fig.~\ref{fig-period}. Let us suppose
each pulse grows with self-similar size distribution,
\begin{equation}
\label{eq-scaling}
N_m(t) = s(t)^a\,F(\xi)\hspace{1.0cm}\mbox{with $\xi = \frac{m}{s(t)}$,}
\end{equation}
where $s(t)$ is a typical size and $a$ is an exponent to be determined.
Substituting Eq.~\eqref{eq-scaling} into Eq.~\eqref{eq-Smol} and balancing
dependences on $t$ requires that $\dot{s} = s^{\nu+\mu+a+2}$. Since the mass
contained in each pulse grows linearly in time, $\int_0^M m\,N_m(t)\,dm = J\,t$.
Substituting Eq.~\eqref{eq-scaling} and differentiating gives $\dot{s}\sim
s^{-a-1}$. Consistency requires
\begin{equation}
\label{eq-exponents}
a=-\frac{\nu+\mu+3}{2} \hspace{1.0cm} s(t) \sim t^\frac{2}{1-\nu-\mu}.
\end{equation}
The period is estimated as the time, $\tau_M$, required for the typical mass,
$s(t)$, to reach $M$. The amplitude, $A_M$, is estimated as the mass supplied in
one period. We thus obtain the following scalings for $\tau_M$ and $A_M$ with
$M$:
\begin{equation}
\label{periodScaling}
\tau_M \sim M^\frac{1-\nu-\mu}{2} \hspace{1.0cm} A_M \sim J\,
M^\frac{1-\nu-\mu}{2}.
\end{equation}
These scalings are verified by the data collapse presented in the inset of
Fig.~\ref{fig-period}. Universality is in a sense restored since the earlier
universal behavior of Eq.\eqref{eq-Cab} can now be understood as the special
case in which $F(\xi)$ has the special form which cancels $s(t)$ from $N_m(t)$
in Eq.\eqref{eq-scaling}.

We believe that the phenomena presented here are unlikely to be restricted to
coagulation. Many driven dissipative systems with conserved currents must
satisfy a locality criterion analogous to the one discussed here \cite{CRZ2007}
and may be candidates for oscillatory behaviour when this criterion is violated.
In particular, the kinetic equation for isotropic 3-wave turbulence, which is
closely analogous to Eq.~\eqref{eq-Smol}, becomes nonlocal when
$\left|\nu-\mu\right|>3$ \cite{CON2009}. Furthermore, the oscillatory behaviour
discussed in this article may even have been already observed experimentally in
measurements of non-equilibrium phase separation of binary mixtures with slowly
ramped temperature \cite{VAV2007,BV2010}. In this system, droplets of one phase
coagulate inside another during demixing with nucleation providing the source of
``monomers'' although the coagulation process is not obviously nonlocal in our
sense. This nevertheless seems like a potentially fruitful direction for
further 
investigation since the theory presented here makes several testable predictions
about the oscillatory kinetics.

\begin{acknowledgments}
C.C. thanks P.~L. Krapivsky for enlightening discussions and encouragement and
acknowledges the financial support of the Engineering and Physical Sciences
Research
Council under grant No. EP/H051295/1.
\end{acknowledgments}

%\bibliography{all}

\begin{thebibliography}{10}%
\makeatletter
\providecommand \@ifxundefined [1]{%
 \ifx #1\undefined \expandafter \@firstoftwo
 \else \expandafter \@secondoftwo
\fi
}%
\providecommand \@ifnum [1]{%
 \ifnum #1\expandafter \@firstoftwo
 \else \expandafter \@secondoftwo
\fi
}%
\providecommand \enquote [1]{``#1''}%
\providecommand \bibnamefont  [1]{#1}%
\providecommand \bibfnamefont [1]{#1}%
\providecommand \citenamefont [1]{#1}%
\providecommand\href[0]{\@sanitize\@href}%
\providecommand\@href[1]{\endgroup\@@startlink{#1}\endgroup\@@href}%
\providecommand\@@href[1]{#1\@@endlink}%
\providecommand \@sanitize [0]{\begingroup\catcode`\&12\catcode`\#12\relax}%
\@ifxundefined \pdfoutput {\@firstoftwo}{%
 \@ifnum{\z@=\pdfoutput}{\@firstoftwo}{\@secondoftwo}%
}{%
 \providecommand\@@startlink[1]{\leavevmode}%
 \providecommand\@@endlink[0]{}%
}{%
 \providecommand\@@startlink[1]{%
  \leavevmode
  \pdfstartlink
   attr{/Border[0 0 1 ]/H/I/C[0 1 1]}%
   user{/Subtype/Link/A<</Type/Action/S/URI/URI(#1)>>}%
  \relax
 }%
 \providecommand\@@endlink[0]{\pdfendlink}%
}%
\providecommand \url  [0]{\begingroup\@sanitize \@url }%
\providecommand \@url [1]{\endgroup\@href {#1}{\urlprefix}}%
\providecommand \urlprefix [0]{URL }%
\providecommand \Eprint[0]{\href }%
\@ifxundefined \urlstyle {%
  \providecommand \doi [1]{doi:\discretionary{}{}{}#1}%
}{%
  \providecommand \doi [0]{doi:\discretionary{}{}{}\begingroup
  \urlstyle{rm}\Url }%
}%
\providecommand \doibase [0]{http://dx.doi.org/}%
\providecommand \Doi[1]{\href{\doibase#1}}%
\providecommand \bibAnnote [3]{%
  \BibitemShut{#1}%
  \begin{quotation}\noindent
    \textsc{Key:}\ #2\\\textsc{Annotation:}\ #3%
  \end{quotation}%
}%
\providecommand \bibAnnoteFile [2]{%
  \IfFileExists{#2}{\bibAnnote {#1} {#2} {\input{#2}}}{}%
}%
\providecommand \typeout [0]{\immediate \write \m@ne }%
\providecommand \selectlanguage [0]{\@gobble}%
\providecommand \bibinfo [0]{\@secondoftwo}%
\providecommand \bibfield [0]{\@secondoftwo}%
\providecommand \translation [1]{[#1]}%
\providecommand \BibitemOpen[0]{}%
\providecommand \bibitemStop [0]{}%
\providecommand \bibitemNoStop [0]{.\EOS\space}%
\providecommand \EOS [0]{\spacefactor3000\relax}%
\providecommand \BibitemShut [1]{\csname bibitem#1\endcsname}%
%</preamble>
\bibitem{LEY2003}%
  \BibitemOpen
  \bibfield{author}{%
  \bibinfo {author} {\bibfnamefont{F.}~\bibnamefont{Leyvraz}},\ }%
  \bibfield{journal}{%
  \bibinfo {journal} {Phys. Reports}\ }%
  \textbf{\bibinfo {volume} {383}},\ \bibinfo {pages} {95} (\bibinfo {month}
  {Aug.}\ \bibinfo {year} {2003})%
  \bibAnnoteFile{NoStop}{LEY2003}%
\bibitem{FFS2002}%
  \BibitemOpen
  \bibfield{author}{%
  \bibinfo {author} {\bibfnamefont{G.}~\bibnamefont{{Falkovich}}}, \bibinfo
  {author} {\bibfnamefont{A.}~\bibnamefont{{Fouxon}}},\ and\ \bibinfo {author}
  {\bibfnamefont{M.~G.}\ \bibnamefont{{Stepanov}}},\ }%
  \bibfield{journal}{%
  \bibinfo {journal} {Nature}\ }%
  \textbf{\bibinfo {volume} {419}},\ \bibinfo {pages} {151} (\bibinfo {year}
  {2002})%
  \bibAnnoteFile{NoStop}{FFS2002}%
\bibitem{KA2011}%
  \BibitemOpen
  \bibfield{author}{%
  \bibinfo {author} {\bibfnamefont{Y.~A.}\ \bibnamefont{{Kryukov}}}\ and\
  \bibinfo {author} {\bibfnamefont{J.~G.}\ \bibnamefont{{Amar}}},\ }%
  \bibfield{journal}{%
  \bibinfo {journal} {Phys. Rev. E}\ }%
  \textbf{\bibinfo {volume} {83}},\ \bibinfo {pages} {041611} (\bibinfo {year}
  {2011})%
  \bibAnnoteFile{NoStop}{KA2011}%
\bibitem{BBK2009}%
  \BibitemOpen
  \bibfield{author}{%
  \bibinfo {author} {\bibfnamefont{N.~V.}\ \bibnamefont{{Brilliantov}}},
  \bibinfo {author} {\bibfnamefont{A.~S.}\ \bibnamefont{{Bodrova}}},\ and\
  \bibinfo {author} {\bibfnamefont{P.~L.}\ \bibnamefont{{Krapivsky}}},\ }%
  \bibfield{journal}{%
  \bibinfo {journal} {J. Stat. Mech.: Theor. E.}\ }%
  \textbf{\bibinfo {volume} {6}},\ \bibinfo {pages} {11} (\bibinfo {year}
  {2009})%
  \bibAnnoteFile{NoStop}{BBK2009}%
\bibitem{WHI1982}%
  \BibitemOpen
  \bibfield{author}{%
  \bibinfo {author} {\bibfnamefont{W.~H.}\ \bibnamefont{{White}}},\ }%
  \bibfield{journal}{%
  \bibinfo {journal} {J. Colloid Interface Sci.}\ }%
  \textbf{\bibinfo {volume} {87}},\ \bibinfo {pages} {204} (\bibinfo {year}
  {1982})%
  \bibAnnoteFile{NoStop}{WHI1982}%
\bibitem{HAK1987}%
  \BibitemOpen
  \bibfield{author}{%
  \bibinfo {author} {\bibfnamefont{H.}~\bibnamefont{Hayakawa}},\ }%
  \bibfield{journal}{%
  \bibinfo {journal} {J. Phys. A}\ }%
  \textbf{\bibinfo {volume} {20}},\ \bibinfo {pages} {L801} (\bibinfo {year}
  {1987})%
  \bibAnnoteFile{NoStop}{HAK1987}%
\bibitem{CRZ2004}%
  \BibitemOpen
  \bibfield{author}{%
  \bibinfo {author} {\bibfnamefont{C.}~\bibnamefont{Connaughton}}, \bibinfo
  {author} {\bibfnamefont{R.}~\bibnamefont{Rajesh}},\ and\ \bibinfo {author}
  {\bibfnamefont{O.}~\bibnamefont{Zaboronski}},\ }%
  \bibfield{journal}{%
  \bibinfo {journal} {Phys. Rev. E}\ }%
  \textbf{\bibinfo {volume} {69}},\ \bibinfo {pages} {061114} (\bibinfo {year}
  {2004})%
  \bibAnnoteFile{NoStop}{CRZ2004}%
\bibitem{KON01}%
  \BibitemOpen
  \bibfield{author}{%
  \bibinfo {author} {\bibfnamefont{V.}~\bibnamefont{Kontorovich}},\ }%
  \bibfield{journal}{%
  \bibinfo {journal} {Physica D}\ }%
  \textbf{\bibinfo {volume} {152--153}},\ \bibinfo {pages} {676} (\bibinfo
  {year} {2001})%
  \bibAnnoteFile{NoStop}{KON01}%
\bibitem{PK1997}%
  \BibitemOpen
  \bibfield{author}{%
  \bibinfo {author} {\bibfnamefont{H.}~\bibnamefont{Pruppacher}}\ and\ \bibinfo
  {author} {\bibfnamefont{J.}~\bibnamefont{Klett}},\ }%
  \emph{\bibinfo {title} {Microphysics of Clouds and Precipitation}},\ \bibinfo
  {edition} {2nd}\ ed.\ (\bibinfo {publisher} {Kluwer Academic Publishers},\
  \bibinfo {address} {Dordrecht, The Netherlands},\ \bibinfo {year} {1997})%
  \bibAnnoteFile{NoStop}{PK1997}%
\bibitem{ZLF92}%
  \BibitemOpen
  \bibfield{author}{%
  \bibinfo {author} {\bibfnamefont{V.}~\bibnamefont{Zakharov}}, \bibinfo
  {author} {\bibfnamefont{V.}~\bibnamefont{Lvov}},\ and\ \bibinfo {author}
  {\bibfnamefont{G.}~\bibnamefont{Falkovich}},\ }%
  \emph{\bibinfo {title} {Kolmogorov Spectra of Turbulence}}\ (\bibinfo
  {publisher} {Springer-Verlag},\ \bibinfo {address} {Berlin},\ \bibinfo {year}
  {1992})%
  \bibAnnoteFile{NoStop}{ZLF92}%
\bibitem{KC2012}%
  \BibitemOpen
  \bibfield{author}{%
  \bibinfo {author} {\bibfnamefont{P.~L.}\ \bibnamefont{{Krapivsky}}}\ and\
  \bibinfo {author} {\bibfnamefont{C.}~\bibnamefont{Connaughton}},\ }%
  \enquote{\bibinfo {title} {{Driven Brownian coagulation of polymers}},}\
  (\bibinfo {year} {2012}),\ \bibinfo {note} {j. Chem. Phys. (to appear)}%
  \bibAnnoteFile{NoStop}{KC2012}%
\bibitem{HNS2008}%
  \BibitemOpen
  \bibfield{author}{%
  \bibinfo {author} {\bibfnamefont{P.}~\bibnamefont{{Horvai}}}, \bibinfo
  {author} {\bibfnamefont{S.~V.}\ \bibnamefont{{Nazarenko}}},\ and\ \bibinfo
  {author} {\bibfnamefont{T.~H.~M.}\ \bibnamefont{{Stein}}},\ }%
  \bibfield{journal}{%
  \bibinfo {journal} {J. Stat. Phys.}\ }%
  \textbf{\bibinfo {volume} {130}},\ \bibinfo {pages} {1177} (\bibinfo {year}
  {2008})%
  \bibAnnoteFile{NoStop}{HNS2008}%
\bibitem{BCSZ2011}%
  \BibitemOpen
  \bibfield{author}{%
  \bibinfo {author} {\bibfnamefont{R.~C.}\ \bibnamefont{{Ball}}}, \bibinfo
  {author} {\bibfnamefont{C.}~\bibnamefont{{Connaughton}}}, \bibinfo {author}
  {\bibfnamefont{T.~H.~M.}\ \bibnamefont{{Stein}}},\ and\ \bibinfo {author}
  {\bibfnamefont{O.}~\bibnamefont{{Zaboronski}}},\ }%
  \bibfield{journal}{%
  \bibinfo {journal} {Phys. Rev. E}\ }%
  \textbf{\bibinfo {volume} {84}},\ \bibinfo {pages} {011111} (\bibinfo {year}
  {2011})%
  \bibAnnoteFile{NoStop}{BCSZ2011}%
\bibitem{LEE2000}%
  \BibitemOpen
  \bibfield{author}{%
  \bibinfo {author} {\bibfnamefont{M.}~\bibnamefont{Lee}},\ }%
  \bibfield{journal}{%
  \bibinfo {journal} {Icarus}\ }%
  \textbf{\bibinfo {volume} {143}},\ \bibinfo {pages} {74} (\bibinfo {year}
  {2000})%
  \bibAnnoteFile{NoStop}{LEE2000}%
\bibitem{LEE2001}%
  \BibitemOpen
  \bibfield{author}{%
  \bibinfo {author} {\bibfnamefont{M.}~\bibnamefont{Lee}},\ }%
  \bibfield{journal}{%
  \bibinfo {journal} {J. Phys. A: Math. Gen.}\ }%
  \textbf{\bibinfo {volume} {34}},\ \bibinfo {pages} {10219} (\bibinfo {year}
  {2001})%
  \bibAnnoteFile{NoStop}{LEE2001}%
\bibitem{MOB2010}%
  \BibitemOpen
  \bibfield{author}{%
  \bibinfo {author} {\bibfnamefont{M.}~\bibnamefont{Mobilia}},\ }%
  \bibfield{journal}{%
  \bibinfo {journal} {J. Theor. Bio.}\ }%
  \textbf{\bibinfo {volume} {264}},\ \bibinfo {pages} {1 } (\bibinfo {year}
  {2010})%
  \bibAnnoteFile{NoStop}{MOB2010}%
\bibitem{ALD1999}%
  \BibitemOpen
  \bibfield{author}{%
  \bibinfo {author} {\bibfnamefont{D.~J.}\ \bibnamefont{Aldous}},\ }%
  \bibfield{journal}{%
  \bibinfo {journal} {Bernoulli}\ }%
  \textbf{\bibinfo {volume} {5}},\ \bibinfo {pages} {3} (\bibinfo {month}
  {Feb.}\ \bibinfo {year} {1999})%
  \bibAnnoteFile{NoStop}{ALD1999}%
\bibitem{CRZ2007}%
  \BibitemOpen
  \bibfield{author}{%
  \bibinfo {author} {\bibfnamefont{C.}~\bibnamefont{Connaughton}}, \bibinfo
  {author} {\bibfnamefont{R.}~\bibnamefont{Rajesh}},\ and\ \bibinfo {author}
  {\bibfnamefont{O.}~\bibnamefont{Zaboronski}},\ }%
  \bibfield{journal}{%
  \bibinfo {journal} {Phys. Rev. Lett.}\ }%
  \textbf{\bibinfo {volume} {98}},\ \bibinfo {pages} {080601} (\bibinfo {year}
  {2007})%
  \bibAnnoteFile{NoStop}{CRZ2007}%
\bibitem{CON2009}%
  \BibitemOpen
  \bibfield{author}{%
  \bibinfo {author} {\bibfnamefont{C.}~\bibnamefont{{Connaughton}}},\ }%
  \bibfield{journal}{%
  \bibinfo {journal} {Physica D}\ }%
  \textbf{\bibinfo {volume} {238}},\ \bibinfo {pages} {2282} (\bibinfo {year}
  {2009})%
  \bibAnnoteFile{NoStop}{CON2009}%
\bibitem{VAV2007}%
  \BibitemOpen
  \bibfield{author}{%
  \bibinfo {author} {\bibfnamefont{J.}~\bibnamefont{Vollmer}}, \bibinfo
  {author} {\bibfnamefont{G.~K.}\ \bibnamefont{Auernhammer}},\ and\ \bibinfo
  {author} {\bibfnamefont{D.}~\bibnamefont{Vollmer}},\ }%
  \bibfield{journal}{%
  \bibinfo {journal} {Phys. Rev. Lett.}\ }%
  \textbf{\bibinfo {volume} {98}},\ \bibinfo {pages} {115701} (\bibinfo {year}
  {2007})%
  \bibAnnoteFile{NoStop}{VAV2007}%
\bibitem{BV2010}%
  \BibitemOpen
  \bibfield{author}{%
  \bibinfo {author} {\bibfnamefont{I.~J.}\ \bibnamefont{{Benczik}}}\ and\
  \bibinfo {author} {\bibfnamefont{J.}~\bibnamefont{{Vollmer}}},\ }%
  \bibfield{journal}{%
  \bibinfo {journal} {Europhys. Lett.}\ }%
  \textbf{\bibinfo {volume} {91}},\ \bibinfo {pages} {36003} (\bibinfo {year}
  {2010})%
  \bibAnnoteFile{NoStop}{BV2010}%
\end{thebibliography}

%Merlin.mbs v4.21 2009-07-09.
%

\appendix

\section{Algorithm for finding exact stationary solution of the Smoluchowski equation}
\label{app-generateStatDist}

\paragraph*{}
Consider the discrete form of the stationary Smoluchowski equation:

\begin{align}
\nonumber  0 &= \frac{1}{2}  \sum_{m_1=1}^{m-1}\, K(m_1,m-m_1)\, N_{m_1}\, N_{m-m_1}\\
\label{eq-stationarySmol}  &- N_m  \sum_{m_1 =1}^{M}\, K(m,m_1)\, N_{m_1} + J_{m,1}.
\end{align}
We use the kernel  considered in the main text:
\begin{equation}
\label{eq-modelKernelApp}
K(m_1, m_2) = \frac{g}{2}\left( m_1^\mu m_2^\nu + m_1^\nu m_2^\mu \right).
\end{equation}
Here $g$ is a constant which can be helpful to keep track of dimensions but which is usually set equal to one.
Denote the $p^\mathrm{th}$ moment of the size distribution by 
\begin{displaymath} 
\mathcal{M}_p = \sum_{m_1=1}^{M} m_1^{p} N_{m_1}.
\end{displaymath}
If the moments $\mathcal{M}_\mu$ and $\mathcal{M}_\nu$ were known, we could find the solution of (\ref{eq-stationarySmol}) 
by iteration:
\begin{equation}
N_m = \frac{ \mathcal{G}_m +  \frac{2 J}{g}\delta_{m,1} }
   { \left( m^{\mu} \mathcal{M}_{\nu}  + m^{\nu} \mathcal{M}_{\mu} \right) },
\label{eq-momentstatSmol}   
\end{equation}
where  $\mathcal{G}_m$ depends only on the densities of clusters with masses less than $m$ and is given by
\begin{align} 
  \mathcal{G}_m = \frac{1}{2} \sum_{m_1=1}^{m-1} K(m_1,m-m_1)\, N_{m_1}\, N_{m-m_1}.
\end{align}
The starting value is obtained by setting $m=1$ which gives the monomer density
\begin{align}
 N_1 &= \frac{2J/g}{\mathcal{M}_{\nu}  +  \mathcal{M}_{\mu}}.
\end{align}

\paragraph*{}
 Given that the solution can be expressed in terms of the two moments $\mathcal{M}_\mu$ and $\mathcal{M}_\nu$, the task is now to self-consistently determine the values of these moments. We can approach this task as a simple two-dimensional optimization problem. Eq. \eqref{eq-momentstatSmol} expresses $N_m$ as a function of the pair of moments $ N_m(\mathcal{M}_{\mu}, \mathcal{M}_{\nu})$.  We create an objective function, $\Psi(\mathcal{M}_{\mu}, \mathcal{M}_{\nu})$, as follows:

\begin{align}
\nonumber \Psi(\mathcal{M}_{\mu}, \mathcal{M}_{\nu}) &= \left[\mathcal{M}_{\mu} - \sum_{m=1}^M m^{\mu} N_m(\mathcal{M}_{\mu}, \mathcal{M}_{\nu})\right]^2  \\
 & \quad + \left[\mathcal{M}_{\nu} - \sum_{m=1}^M m^{\nu} N_m(\mathcal{M}_{\mu}, \mathcal{M}_{\nu})\right]^2.
 \end{align}
The correct values of $\mathcal{M}_\mu$ and $\mathcal{M}_\nu$, which we denote by $\mathcal{M}_{\mu}*$ and $\mathcal{M}_{\nu}*$, can be found by minimising $\Psi(\mathcal{M}_{\mu}, \mathcal{M}_{\nu})$:
\begin{align}
(\mathcal{M}_{\mu}*, \mathcal{M}_{\nu}*) &= \arg \min_{(\mathcal{M}_{\mu}, \mathcal{M}_{\nu})} \Psi(\mathcal{M}_{\mu}, \mathcal{M}_{\nu}).
\label{eq-statgenminimise}
\end{align}

This can be done with any numerical minimization algorithm. We used the Nelder-Mead downhill simplex method. The solution thus obtained is exact to within computational error since no approximations have been made in formulating this procedure. 

We remark that the problem does not have to be formulated as an optimization problem. One could treat it as two-dimensional root-finding problem. Furthermore, by summing Eq.\eqref{eq-stationarySmol}, one can derive an independent relationship between $\mathcal{M}_\mu$ and $\mathcal{M}_\nu$ (in the limit $M\to \infty$) which further reduces the problem to a one-dimensional root finding problem. We did experiment with some of these alternatives but settled on the procedure described above as the most numerically stable and reliable approach.

\section{Derivation of the nonlocal Smoluchowski equation}
\label{app-differentialApprox}

We assume without loss of generality that $\nu\geq\mu$. It is the
combination $\gamma=\nu-\mu-1$ which determines the locality of the 
stationary solution of Smoluchowski's equation. The nonlocal case
corresponds to $\gamma >0$.  We can use the differential approximation outlined
in \cite{HNS2008} to describe the stationary state in this regime. The first step is to
rewrite the Smoluchowski equation in a particular form.  Terms describing interactions 
between a reference mass, $m$, and masses less
than $\frac{m}{2}$ are gathered together in one group. Those describing
interactions with masses larger than $\frac{m}{2}$ are gathered together
in a second group. Splitting the integrals appropriately and performing
some manipulations we obtain:
\begin{eqnarray}
\nonumber \dot{N}_m &=& \!\int_0^\frac{m}{2}\!\!\!\!\!dm_1 \left[ K(m_1,m\! -\! m_1) N_{m-m_1} \!\! - K(m_1,m) N_{m}\right] N_{m_1}\\
\label{eq-Smol2} &-& N_m  \int_\frac{m}{2}^{M}dm_1 K(m,m_1) N_{m_1}  + J\ \delta(m-m_0).
\end{eqnarray}
Consider the first term which accounts for all interactions between clusters
of mass $m$ and those having mass, $m_1 <\frac{m}{2}$. If the cascade is
nonlocal, these interactions are primarily with those clusters having
$m_1 \ll \frac{m}{2}$, in which case the integrand is strongly concentrated
in the region $m_1 \ll \frac{m}{2}$. We can then Taylor expand with
respect to $m_1$ and neglect all terms of $O(m_1^2)$ or higher to obtain:
\begin{eqnarray}
\nonumber \partial_t N_m &=&-\pd{}{m}\left[\int_0^\frac{m}{2} dm_1 K(m,m_1)\,N_{m_1}\,N_m \right]\\
\label{eq-Smol3} &-& N_m  \int_\frac{m}{2}^{M}dm_1 K(m,m_1) N_{m_1} \\
 \nonumber &+& J\ \delta(m-m_0).
\end{eqnarray}
With the kernel given by Eq.~(\ref{eq-modelKernelApp}), we get the following
equation for the stationary state:
\begin{eqnarray}
\nonumber 0 &=&-\dd{}{m}\left[\mathcal{M}_{\nu+1}^< m^\mu N_m \right] -\dd{}{m}\left[\mathcal{M}_{\mu+1}^< m^\nu N_m \right]\\
\label{eq-Smol4} &-& \mathcal{M}_\nu^> m^\mu N_m - \mathcal{M}_\mu^> m^\nu N_m + \frac{2\,J}{g}\ \delta(m-m_0),
\end{eqnarray}
where $\mathcal{M}^>_p$ and $\mathcal{M}^<_p$ denote the upper and lower partial moments:
\begin{eqnarray}
\mathcal{M}_p^< &=& \int_{m_0}^\frac{m}{2} dm_1 m_1^p N_{m_1}\\
\mathcal{M}_p^> &=& \int_\frac{m}{2}^M dm_1 m_1^p N_{m_1}.
\end{eqnarray}
The dominant terms in this equation when $m\gg m_0$ will turn out to be
\begin{eqnarray}
\nonumber 0 &=&-\dd{}{m}\left[\mathcal{M}_{\mu+1}^< m^\nu N_m \right]- \mathcal{M}_\nu^> m^\mu N_m \\
\nonumber &+& \frac{2\,J}{g}\ \delta(m-m_0).
\end{eqnarray}
This statement will have to be justified a-posteriori. In order to make further
progress let us assume that the error made by extending the upper and lower
limits of integration in the partial moments $\mathcal{M}_{\mu+1}^<$ and
$\mathcal{M}_\nu$ to $M$ and $m_0$ respectively is small. This will also
have to be justified a-posteriori. The resulting equation is:
\begin{eqnarray}
\label{eq-nonlocalSmolEqApp} 0 &=&-\dd{}{m}\left[\mathcal{M}_{\mu+1} m^\nu N_m \right]- \mathcal{M}_\nu m^\mu N_m \\
\nonumber &+& \frac{2\,J}{g}\ \delta(m-m_0).
\end{eqnarray}
This is Eq.(5) in the main text.

\section{Asypmtotic solution of nonlocal Smoluchowski equation}
\label{app-nonlocalSolution}

Eq.\eqref{eq-nonlocalSmolEqApp} is a linear equation and can be readily integrated to give
\begin{equation}
\label{eq-nonlocalSolnApp}
N_m = C\,e^{\frac{\beta}{\gamma}m^{-\gamma}}\,m^{-\nu}
\end{equation}
where $\beta$ is a ratio of moments
\begin{equation}
\label{eq-consistency}
\beta = \frac{\mathcal{M}_\nu}{\mathcal{M}_{\mu+1}}
\end{equation}
and $C$ is a constant of integration. The non-trivial aspect of the problem is that the moments $\mathcal{M}_\nu$ and $\mathcal{M}_{\mu+1}$ must be determined self-consistently from this solution. This cannot be done analytically but an
asymptotic solution for large cutoff, $M$, can be found which we now describe. A general moment of order $\alpha$ is
\begin{eqnarray}
\nonumber \mathcal{M}_\alpha &=& C\int_{m_0}^M d m \ m^{\alpha-\nu} e^{\frac{\beta}{\gamma}m^{-\gamma}}\\
\label{eq-Malpha} &=& \frac{C}{\beta} \left(\frac{\beta}{\gamma}\right)^{\zeta_\alpha+1}\, \int_{\frac{\beta}{\gamma}\,M^{-\gamma}}^{\frac{\beta}{\gamma}\,m_0^{-\gamma}} t^{-\zeta_\alpha -1} e^t dt
\end{eqnarray}
where we have introduced the shorthand notation $\zeta_\alpha$ for the combination
\begin{displaymath}
\zeta_\alpha = \frac{1-\nu+\alpha}{\gamma}.
\end{displaymath}
Since $\mu>\nu+1$ in the nonlocal regime, we would expect the moment $\mathcal{M}_\nu$ to grow faster than 
$\mathcal{M}_{\mu+1}$ as the cutoff, $M$, is increased. We therefore expect $\beta$ to grow as $M$ grows. Let us 
suppose that it does not grow faster than $M^\gamma$. If this is the case then the upper limit of the integral in
Eq.\eqref{eq-Malpha} tends to infinity as $M$ grows while the lower limit tends to zero. We are therefore interested in
the behaviour of the integral
\begin{displaymath}
I(\epsilon, \Lambda,\zeta_\alpha) \int_\epsilon^\Lambda t^{-\zeta_\alpha -1} e^t dt
\end{displaymath}
as $\epsilon\to 0$ and $\Lambda\to\infty$. This integral clearly diverges at its upper limit regardless of the value of
$\zeta_\alpha$. The leading order behaviour as the upper limit grows is
\begin{equation}
I(\epsilon, \Lambda,\zeta_\alpha) \sim \Lambda^{-\zeta_\alpha-1}\,e^\Lambda\hspace{1.0cm}\mbox{as $\Lambda\to\infty$}.
\end{equation}
It is divergent at its lower limit if $\zeta_\alpha>0$. The leading order behaviour as the lower limit goes to zero is
\begin{equation}
I(\epsilon, \Lambda,\zeta_\alpha) \sim \frac{1}{\zeta_\alpha} \,\epsilon^{-\zeta_\alpha} \hspace{1.0cm}\mbox{as $\epsilon\to 0$}.
\end{equation}
The moments of immediate interest correspond to $\alpha=\nu$ and $\alpha=\mu+1$ which give values for $\zeta_\alpha$ of 
$\frac{1}{\gamma}$ and $\frac{1-\gamma}{\gamma}$ respectively. For $\alpha=\nu$, e always get a divergence at the lower limit.
For $\alpha=\mu+1$ we get a divergence at the lower limit for $\gamma$ in the range $0<\gamma <1$. With 
this knowledge in mind, our task is now to substitute Eq.\eqref{eq-Malpha} into the consistency condition, 
Eq.\eqref{eq-consistency} and attempt to balance the divergences as $M$ (and thus $\beta$) tends to infinity. One solution
is to balance the divergence coming from the lower limit of $\mathcal{M}_{\mu+1}$ with the one coming from the
upper limit of $\mathcal{M}_{\nu}$. This gives, after some work
\begin{equation}
\label{eq-beta}
\beta \sim \gamma\, m_0^\gamma\, \ln \frac{M}{m_0} \hspace{1.0cm}\mbox{as $M \to\infty$},
\end{equation}
which is consistent with our assumption that $\beta$ should grow with $M$ but slower than $M^\gamma$. We can now
substitute this value for $\beta$ into Eq.\eqref{eq-Malpha} and obtain the leading order behaviour of $\mathcal{M}_{\nu}$
and $\mathcal{M}_{\mu+1}$. We find that $\mathcal{M}_{\nu}$ is dominated by its upper limit and grows as
\begin{equation}
\label{eq-Mnu}
\mathcal{M}_{\nu} \sim C m_0 \left(\frac{M}{m_0}\right) \hspace{1.0cm}\mbox{as $M \to\infty$}.
\end{equation}
On the other hand $\mathcal{M}_{\mu+1}$ is dominated by its lower limit and grows as
\begin{equation}
\label{eq-MmuPlus1}
\mathcal{M}_{\mu+1} \sim C m_0^{1-\gamma}  \left(\frac{M}{m_0}\right)\,\ln  \left(\frac{M}{m_0}\right)^{-1}\hspace{1.0cm}\mbox{as $M \to\infty$}.
\end{equation}
These estimates justify our replacement of the partial moments with full moments in the derivation of Eq.\eqref{eq-nonlocalSmolEqApp}. 

It remains to find the constant $C$. This can be done by requiring that the total mass flux leaving the system is
equal to the input flux, $J$:
\begin{equation}
J = \int_{m_0}^M dm\, m \int_{M-m}^M dm_1 K(m,m_1)\,N_m\,N_{m_1}.
\end{equation}
Substituting the kernel Eq.\eqref{eq-modelKernelApp} into this gives two terms:
\begin{eqnarray*}
\frac{2\,J}{g} &=& \int_{m_0}^M d m\, m^{\mu+1}\,N_m\,\int_{M-m}^M d m_1 m_1^\nu\,N_{m_1}\\
&+& \int_{m_0}^M d m\, m^{\nu+1}\,N_m\,\int_{M-m}^M d m_1 m_1^\mu\,N_{m_1}.
\end{eqnarray*}
With some further analysis one finds that the second term is much smaller than the first term as $M$ grows. Extending the
regions of integration of the partial moments as before we obtain the estimate:
\begin{displaymath}
\frac{2\,J}{g} \sim \mathcal{M}_{\mu+1}\,\mathcal{M}_\nu \hspace{1.0cm}\mbox{as $M \to\infty$}.
\end{displaymath}
Using Eqs. \eqref{eq-Mnu} and \eqref{eq-MmuPlus1} we obtain
\begin{equation}
C = \frac{m_0}{M} \sqrt{\frac{2\,\gamma\,J}{g}\, m_0^{\gamma-2}\,\ln\left(\frac{M}{m_0}\right)}.
\end{equation}
Putting this together with Eqs. \eqref{eq-nonlocalSolnApp} and \eqref{eq-beta} we finally obtain:
\begin{eqnarray}
\label{eq-nonlocalSSApp}
N_m^* &\sim& \sqrt{\frac{2\,\gamma\,J}{g}} m_0^{-\frac{\nu+\mu+3}{2}} \sqrt{\ln\left(\frac{M}{m_0}\right)} \left(\frac{M}{m_0}\right)^{-1}\\
\nonumber & &\times \left(\frac{M}{m_0}\right)^{\left(\frac{m}{m_0}\right)^{-\gamma}} \left(\frac{m}{m_0}\right)^{-\nu}.
\end{eqnarray}
The explicit dependence on the monomer mass, $m_0$ (which we usually take
equal to 1) has been retained in order to make the dimensional correctness
of the formula clear. Setting $m_0=1$ gives Eq.~(6) in the main text.

\section{Comment on the accompanying movie}
\label{app-movie}

The movie accompanying the arxiv version of this paper shows the time evolution of the density contrast, $N(m,t)/N^*(m)$, relative to the stationary state as a function of cluster size, $m$, in the oscillatory regime. This quantity would be 1 if the stationary state were stable. Both axes are linear. The movie was generated by solving the discrete Smoluchowski equation (without coarsegraining) with $\nu=-\mu = \frac{3}{2}$, a monomer input rate of $J=1$ and a cut-off of $M=100$.  

\end{document}